\documentclass[useAMS,usenatbib]{mn2e}

\usepackage{graphicx}
\usepackage{verbatim}
\usepackage{bm}

\title[High S/N observations and the limits of pulsar timing]{The ultimate limits of precision pulsar timing}

\author[S. Os{\l}owski et al.]{S. Os{\l}owski$^{1,2}$\thanks{E-mail:soslowski@astro.swin.edu.au}, W. van Straten$^{1}$, G. B. Hobbs$^{2}$, M. Bailes$^{1}$ and P. Demorest$^{3}$\\
$^{1}$Swinburne University of Technology, Centre for Astrophysics and Supercomputing, Mail H39, PO Box 218, VIC 3122, Australia\\
$^{2}$CSIRO Astronomy and Space Sciences, Australia Telescope National Facility, P.O. Box 76, Epping, NSW 1710, Australia\\
$^{3}$National Radio Astronomy Observatory, 520 Edgemont Road, Charlottesville, Virginia 22093, USA\\}

\begin{document}

\newcommand{\msun}{\mbox{M$_{\odot}$}}
\newcommand{\rsun}{\mbox{R$_{\odot}$}}
\newcommand{\aopx}{\mbox{$\Delta_{\pi \rm M}$}}
\newcommand{\shap}{\mbox{$\Delta_{\rm S}$}}
\newcommand{\psr}{\mbox{PSR J0437$-$4715}}

\newcommand{\pb}{\mbox{$P_{\rm b}$}}
\newcommand{\pbdot}{\mbox{$\dot{\pb}$}}

\newcommand{\wvect[1]}{{\bsf #1}}
\newcommand{\bvect[1]}{{\bsf #1}$_{\bm 0}$}

\let\oldhat\hat
\renewcommand{\vec}[1]{\mathbf{#1}}
\renewcommand{\hat}[1]{\oldhat{\mathbf{#1}}}

\date{Accepted . Received ; in original form}

\pagerange{\pageref{firstpage}--\pageref{lastpage}} 
\pubyear{2011}

\maketitle

\label{firstpage}
  
\begin{abstract}

We demonstrate that the sensitivity of high-precision pulsar timing experiments
will be ultimately limited by the
broadband intensity modulation that is intrinsic to the pulsar's stochastic
radio signal.  That is, as the peak flux of the pulsar approaches that of the
system equivalent flux density, neither greater antenna gain nor increased
instrumental bandwidth will improve timing precision.  These conclusions
proceed from an analysis of the covariance matrix used to characterise residual
pulse profile fluctuations following the template matching procedure for
arrival time estimation.  We perform such an analysis on 25 hours of
high-precision timing observations of the closest and brightest millisecond
pulsar, \psr.  In these data, the standard deviation of the post-fit arrival
time residuals is approximately four times greater than that predicted by
considering the system equivalent flux density, mean pulsar flux and the
effective width of the pulsed emission.  We develop a technique based on
principal component analysis to mitigate the effects of shape variations on
arrival time estimation and demonstrate its validity using a number of
illustrative simulations. When applied to our observations, the method reduces
arrival time residual noise by approximately 20\%.  We conclude that, owing
primarily to the intrinsic variability of the radio emission from \psr\ at
20\,cm, timing precision in this observing band better than 30 - 40 ns in one
hour is highly unlikely, regardless of future improvements in antenna gain or
instrumental bandwidth.  We describe the intrinsic variability of the pulsar
signal as stochastic wideband impulse modulated self-noise (SWIMS) and argue
that SWIMS will likely limit the timing precision of every millisecond pulsar
currently observed by Pulsar Timing Array projects as larger and more sensitive
antennas are built in the coming decades.

\end{abstract}

\begin{keywords}
pulsars: general -- pulsars: individual (\psr)

\end{keywords}

\section{Introduction}
\label{Section::intro}

The most fundamental property of radio pulsars is their periodic series of
radio pulses that enable their discovery and a myriad of timing applications.
A sub-class of pulsars, known as the millisecond and recycled pulsars, have
spin periods between 1.4 and a few tens of ms and typical spin-down rates of
$\dot{P}\sim10^{-20}$.  Their short periods and low braking torques make them
especially good clocks, and these pulsars exhibit the highest timing precision
\citep{1997A&A...326..924M}. For most of these pulsars a
simple model of the pulsar spin-down, astrometric and orbital parameters can be
determined, enabling the mean time-of-arrival (ToA) of pulses to be predicted
accurately and precisely.  These can be used for several applications, such as
testing the general theory of relativity
\citep[e.g.][]{1982ApJ...253..908T,2006Sci...314...97K}, detecting
irregularities in terestrial time standards \citep{1996A&A...308..290P,
2008MNRAS.387.1583R,2010arXiv1011.5285H} and to attempt the first direct
detection of a stochastic background of gravitational waves \citep[see,
e.g.][]{1983ApJ...265L..39H,1990ApJ...361..300F,2007PhDT........14D,2010MNRAS.407..669Y,2011MNRAS.414.3117V}.
The wealth of information already derived from the precision timing of millisecond
radio pulsars has led many authors to predict the kind of pulsar timing science
possible with the Square Kilometre Array (SKA) by linearly extrapolating
current telescope sensitivities to that of the SKA.

The closest and brightest millisecond pulsar known, \psr\
\citep{1993Natur.361..613J} has been studied by numerous authors with steadily
improving instrumentation.  \citet{1997ApJ...478L..95S} observed the pulsar
using an autocorrelation spectrometer with 128 MHz of bandwidth, and could
model pulse arrival times over two years with a post-fit residual \citep[the
difference between observed and predicted arrival times after fitting for the
pulsar spin, astrometric and binary parameters, etc.;][]{1992RSPTA.341..117T}
standard deviation of $500$\,ns. Noting that the formal uncertainty of arrival
time estimates was typically around 50\,ns, the authors concluded that their
results were limited by polarimetric calibration errors.
\citet{2000ApJ...532.1240B} first proposed the use of the Stokes invariant
interval to mitigate the problems caused by polarisation calibration. This was
later implemented by \citet{2001Natur.412..158V}, who used a combination of
16\,MHz and 20\,MHz baseband recording systems, typical integrations of 1 hour
duration, and coherent dedispersion to obtain a root-mean-square (rms) timing
residual of 130\,ns over 3.4\,yr.  Using new and improved calibration methods
developed by \citet{2004ApJS..152..129V,2006ApJ...642.1004V} and a new baseband
recording and processing system
\citep[CPSR2;][]{2003ASPC..302...57B,2005PhD.....H} with 128 MHz of bandwidth,
\citet{2008ApJ...679..675V} achieved $199$\,ns over 10 years.

None of the above studies achieve the timing precision predicted by the formal
uncertainty in arrival time estimates.  When observing \psr\ in the 20\,cm band
at the Parkes $64$m observatory, the expected rms timing residual from a 256
MHz band with $21$~K system temperature is about 10 and 80\,ns in one hour and
one minute of integration, respectively.  These uncertainties are derived from
the template-matching method used for pulsar timing, in which each observation
of the average pulse profile\footnote{throughout the paper we refer to the
observed averaged phase resolved light curve of the pulsar as the
pulse profile or pulse shape} $O(t)$ is modelled as a scaled (A) and offset (B)
template $S(t)$, rotated by some phase shift $\phi$, with additional white
noise $N(t)$ \citep{1992RSPTA.341..117T,2010IAUS..261..212B}.
\begin{equation}
O(t) = A S(t-\phi) + B + N(t).
\label{Equation::timing}
\end{equation}
It is generally assumed that the summation of many hundreds or thousands of
pulses leads to a stable pulse profile that is characteristic of the pulsar
\citep{1975ApJ...198..661H}. Consideration of only additional white noise,
$N(t)$, in the above equation is equivalent to assuming that the system
equivalent flux density is the only significant source of noise. However for
bright sources and/or high gain antenna, this assumption is no longer tenable
in at least two circumstances. Firstly, when the flux density of the pulsar
approaches the system equivalent flux density (SEFD) of the receiver additional
noise proportional to the pulsar's flux density becomes significant
\citep[e.g.][]{1989AJ.....98.1112K,2001ApJ...554.1197G,2004PASP..116...84G,
2006PASP..118..461G,2009ApJ...694.1413V,2011ApJ...733...51G,
2011ApJ...733...52G}. Secondly, it is known that each single pulse can have
very different morphology and can occur at different pulse phase
\citep{1968Sci...160..758D,1975ApJ...198..661H,1998ApJ...498..365J}. Even
after integrating over many pulse periods, this subpulse
modulation\footnote{The term ``phase jitter''
\citep[e.g.][]{2010arXiv1010.3785C} is sometimes used to describe this
phenomenon but we find it somewhat misleading as it is not only the pulse phase
that varies.} can introduce detectable fluctuations in the average profile
shape and thereby contribute additional noise to timing data. We discuss the
noise balance in more detail in \S \ref{Section::selfnoise}, and argue that
these two contributions should be considered together as they are related and
are described by the same statistical model.

We note that the presence and importance of pulse profile variability have been
discussed in many different contexts.  Some ``classical'' pulsars have been
observed to change between two or more stable profiles -- a phenomenon known as
mode changing --  on time-scales of minutes to hours
\citep[e.g.][]{1970Natur.228.1297B,1982ApJ...258..776B}.  On longer time
scales, pulsars have been discovered whose emission completely switches off for
many days, weeks or even months
\citep[e.g.][]{1979MNRAS.186P..39D,2006Sci...312..549K}.  Recently,
\citet{2010Sci...329..408L} have shown that the pulse profiles for many pulsars
switch between two unique states on time-scales of months to years.
\citet{2011MNRAS.415..251K} recently detected a transient component in
\mbox{PSR J0738$-$4042}, varying on time-scale of years or decades.

The connection between the pulse shape changes and timing noise was made
soon after the discovery of timing noise in the pulsar observations by
\citet{1972ApJ...175..217B}. They studied optical timing observations of the
Crab pulsar and discovered a noise component in the timing residuals which was
well modelled as a random walk in the pulsar spin frequency. The authors also
considered a random walk in the pulse phase and spin frequency derivative, but
found no evidence of such noise in their data. This analysis was extended in a
series of papers
\citep{1975ApJS...29..443G,1975ApJS...29..453G,1975ApJS...29..431G}. The author
presented an analysis method suitable for studies of data with inherent timing
noise. This improved methodology led to the conclusion that noise in the Crab
pulsar timing is dominated by a random walk in the spin frequency but a random
walk in pulsar phase might also be present. In the meantime,
\citet{1974ApJ...191L..63M} described timing noise for two slow pulsars from
radio observations. A few years later, another series of papers
\citep{1980ApJ...237..206H,1980ApJ...237..216C,1980ApJ...239..640C} presented
statistics of timing noise for 37 bright pulsars and concluded that it is a
ubiquitous phenomenon. These authors presented a careful analytical description
of random walks in pulsar phase, spin frequency and its derivative and the
uncertainties in the estimation of their parameters. The last paper in
the series pointed out that the random walk in the pulsar phase can be due to
the random pulse shape changes but concluded that it was unlikely to be the
dominant source of timing noise in the available data. A different dataset was
analysed in a similar manner by \citet{1985ApJS...59..343C} who stated that
either excessive jitter or pulse shape changes are important for a significant
fraction of their sample. They also pointed out that the pulse shape changes
are likely to be universal but their importance varies from object to object.
Later, \citet{1993ASPC...36...43C} detected pulse shape variability in 11 out
of 14 studied objects. These variations were consistent with being caused by
summation of a finite number of pulses. A year later
\citet{1994ApJ...428..713K} studied two millisecond radio pulsars and
discovered timing noise in one of them. The general continuity of properties
between classical and millisecond pulsars suggests that pulse shape changes may
be common in millisecond pulsars as well.

The profile variability of millisecond pulsars (MSPs) has been studied in
relatively few cases.  The single pulses from \mbox{PSR J1939$+$2134} show no
subpulse structure over selected ranges of pulse longitude
\citep{2001ApJ...546..394J} but emit giant pulses as much as 300 times brighter
than the average pulse, that are narrower and systematically delayed with
respect to the main and interpulse components
\citep{1996ApJ...457L..81C,2000ApJ...535..365K}. Several other groups have
argued that MSPs exhibit profile shape changes. Some are associated with
different viewing geometries or with gravitational spin-orbit coupling; e.g.
\mbox{PSR B1913$+$16} \citep{1989ApJ...347.1030W,1998ApJ...509..856K} and
\mbox{PSR B1534$+$12} \citep{1995PhDT.........3A,2000ASPC..202..121S}.
\citet{1997AJ....114.1539B} claimed erratic emission modes from
PSR~\mbox{B1821$-$24}\ but at only one observing frequency.  In another work,
\citet{1999ApJ...520..324K} studied \mbox{PSRs J1022$+$1001} and
\mbox{J1730$-$2304}. In both cases, they detected profile variations over
time-scales of the order of $10$ to $15$ minutes; however the data quality for
the latter did not allow a rigourous statistical analysis.  On the other hand,
\citet{2004MNRAS.355..941H} detected no significant variations in the pulse
profile of \mbox{PSR J1022$+$1001} and demonstrated that the reported profile
shape variations could be explained by polarisation calibration errors.

Small profile changes in \psr\ were described by \citet{1998ApJ...501..823V}
using observations performed at a very low frequency with only a single
polarisation. This result was contested by \citet{1997ApJ...478L..95S}, who
argued that calibration errors were the origin.  \citet{2001MNRAS.326L..33V}
later argued that the variations are intrinsic to the pulsar and correlated
with spiky emission in the varying component.  Variations in the central region
of the profile were also reported by \citet{1997ApJ...486.1019N} with $24$
minute integrations at $428$~MHz but they were not investigated in detail.
\citet{2011MNRAS.tmp.1442L} developed a sharpness statistic but found it
insesitive to profile changes in \psr.

As described in more detail in \S \ref{Section::selfnoise}, this work focuses
on the stochastic fluctuations in total intensity that arise from the subpulse
structure observed in single pulses and their effect on the timing precision
attainable for \psr.  In \S \ref{Section::observations} our observations are
described along with the applied data processing followed by results of timing
our observations. In \S \ref{Section::method} we describe a statistical method
useful for detecting profile shape variations, then apply it to simulated data
as a demonstration of how it can be used to correct ToA residuals. The results
of the statistical analysis are presented in \S \ref {Section::results}. We
summarise our findings and discuss their consequences in \S
\ref{Section::discussion}, which also contains a discussion of other possible
problems that prevent us from reaching the theoretical timing accuracy. Finally
we draw our conclusions in \S \ref{Section::conclusions}.

\section {Stochastic Wideband Impulse Modulated Self-Noise}
\label{Section::selfnoise}

The noise $N(t)$ in equation \ref{Equation::timing} is normally assumed to be
dominated by the white radiometer noise. In practice, for bright sources and/or
high gain antennas, two additional effects may contribute significantly. In
this section we discuss the noise balance for pulsars and demonstrate that the
different contributions are well described by a single statistical model.

Firstly, the noise balance has to include the source itself when the flux density
of the pulsar approaches the SEFD. This noise is intrinsic to the source and is
accordingly called ``self-noise''. In the case of \psr, the mean flux at the peak of the
pulse profile is of the order of 5 Jy; this contributes only $\sim 2\%$ to the
standard deviation of the total intensity, which is dominated by the SEFD $\sim
27$\,Jy of the  20\,cm multibeam receiver \citep{2001MNRAS.328...17M} commonly
used for pulsar timing observations at Parkes.  Any source that can be
described as noise (e.g. thermal emission) will contribute to the variance of
the observed total intensity of the source.  When the signal-to-noise ratio
($S/N$) is low, this contribution is negligible. We note that throughout the
paper we use $S/N$ values calculated using the noise measured in the off-pulse
baseline.

Secondly, dramatic subpulse amplitude modulation is a ubiquitous feature of
radio pulsar emission \citep{1975ApJ...197..185R} that spans orders of
magnitude in intensity and duration and is a broadband phenomenon
\citep[e.g.][]{1968Sci...162.1481S,1975ApJ...195..513T,1975ApJ...196...83M,1993ApJ...417..735H,2007ApJ...670..693H,2007MNRAS.377.1383W}.
The subpulse emission from \psr\ is well studied; \cite{1998ApJ...498..365J}
observe an exponential distribution of peak subpulse intensities with a mean
flux density of 16.6 Jy, which is comparable to the SEFD. More importantly, the
mean subpulse width of $65$\,$\mu$s \citep{1998ApJ...498..365J} is about an
order of magnitude larger than the interval required to sample the mean pulse
profile of \psr.  Consequently, subpulse intensity fluctuations introduce
detectable variations in the average pulse profile. Given the stochastic nature
of subpulse structure, these fluctuations in mean pulse profile shape can be
expected to introduce significant additional noise in derived arrival time
estimates.

In one minute, \psr\ turns $\sim 10^4$ times and emits at least $\sim
2\times10^3$ subpulses \citep{1998ApJ...498..365J}. After integrating over
such a large number of emission events, it is no longer practical to consider
the impact of an individual subpulse.  Rather, it becomes necessary to
describe the effects of subpulse modulation in purely statistical terms using
the fourth moments of the electric field \citep{1975ApJ...197..185R}.  From
this perspective, the subpulse modulation phenomenon is a noise process that
contributes to the autocorrelation of the total intensity
\citep{1975ApJ...197..185R}. For a given source flux density, amplitude
modulation increases the variance of the total intensity and, depending on the
time-scale of the modulations and the sampling interval of the instrument,
introduces power at non-zero delays in the autocorrelation of the total
intensity.

Measured statistical distributions of subpulse intensities vary between
sources and as a function of pulse longitude \citep[e.g. power law, log normal,
etc.; for an excellent review, see][]{2004ApJ...610..948C}.  Regardless of the
original distribution, after a large number of pulses have been integrated, the
central limit theorem applies and profile shape variations are well described
by a multivariate normal distribution. The covariance matrix that quantifies
this distribution contains phase-resolved information about the mean
autocorrelation of the total intensity. 

To summarise, depending on the pulsar's flux density, its emission properties
and the used instrument, we can distinguish three noise regimes:
\begin{enumerate}

\item {\it First regime:} The pulsar's flux density is much smaller than the
SEFD of the instrument. This is the classic regime, in which the noise is
temporally uncorrelated between the phase bins and homoscedastic (i.e., the
variance of noise is the same in each phase bin). In this regime the covariance
matrix of the pulse profiles is well approximated by a diagonal matrix with all
the elements on the diagonal equal to the variance of SEFD.

\item {\it Second regime:} The pulsar's flux density approaches or exceeds the
SEFD of the received used. In this regime the self-noise cannot be neglected.
The noise is still temporally uncorrelated between the phase bins, but it is
heteroscedastic; that is the variance of the noise is different in each bin and
proportional to the sum of squares of the pulsar's flux density and the SEFD.
The covariance matrix of the data is still diagonal, but the non-zero elements
are no longer equal. In this regime, the on-pulse noise is no longer measured
by the off-pulse noise and using the latter to calculate the $S/N$ can lead to
overestimating the achievable timing precision.

\item {\it Third regime:} The pulsar is heavily amplitude modulated with the
modulated flux approaching or exceeding the SEFD of the instrument. Even though
the self-noise contribution may be negligible, the modulated subpulses can
approach the SEFD of the receiver, thus contributing significant `noise' to
the averaged pulse profile. If the sampling rate is high enough to resolve the
subpulse structure, the noise in different phase bins will be heteroscedastic
and temporally correlated. The off-diagonal elements of the covariance matrix
will be non-zero in this regime. If the subpulses are not resolved, amplitude
modulation may still be evident in the variation of the modulation index as a
function of pulse phase.  The broadband nature of the impulses will also lead
to spectral correlation of the noise, which can be detected by measuring the
covariance of intensity fluctuations in different frequency channels.
Therefore, when analysing only the covariance matrix SWIMS might be confused
with self-noise.

\end{enumerate}

In this paper, we investigate the effects of the third regime, which we call
stochastic wideband impulse modulated self-noise (SWIMS), on pulse arrival time
estimation.   This pulsar-intrinsic noise has also been called pulse-phase
jitter (or jitter noise), ``intermittent emission'' \citep{2011ApJ...733...52G}
or simply self-noise.  We will demonstrate that the timing precision of \psr\
is currently limited by SWIMS and that its effect on the mean pulse profile is
readily detectable.

As a function of integration length $T$, the covariance matrix scales as
$T^{-1}$; that is, the effects of SWIMS are reduced by integration, regardless
of the dominating source of noise. The fact that the covariance matrix scales
as $T^{-1}$ allows us to study the statistics of single pulses with longer
integrations.  If any pulse-to-pulse correlation were present in data, such as
arising from drifting subpulses, nulling, mode changing, scattering or
polarisation calibration errors, the scaling of the covariance matrix with
integration length would deviate from the above proportionality. We note that
the relative contribution of source-intrinsic noise to the covariance matrix
will vary as the flux density of the pulsar varies, primarily owing to
interstellar scintillation.  However, after averaging over many scintillation
time scales, the relative contribution of SWIMS compared to the SEFD is
constant \citep{1989AJ.....98.1112K}; therefore, the relative importance of
SWIMS is independent of integration length. We note that in the first or second
regime,  noise can be reduced by increasing the bandwidth. However, because the
intensity fluctuations are typically correlated over wide bandwidths, the noise
due to subpulse modulation is not reduced by increasing the bandwidth and only
longer integration times and active mitigation can improve timing precision in
the third regime.

\citet{2011ApJ...733...52G} recently performed a detailed analysis of
impulse-modulated self-noise in the context of interstellar scintillation
observations and concluded that self-noise may limit pulsar timing precision.
In this paper, we explore the impact of both temporal and spectral correlations
of intensity fluctuations on pulsar timing and consider active mitigation of
SWIMS.

\section {Observations and data processing}
\label{Section::observations}

Observations of \psr\ were recorded during one week of February 2010 using the
Parkes 64\,m radio telescope and the central beam of the 20\,cm multibeam
receiver \citep{1996PASA...13..243S}.  The third generation of the Pulsar
Digital Filterbank (PDFB3) digitised the voltage data from two orthogonally
polarised 256 MHz bands and formed 1024 frequency channels using a polyphase
filterbank.  After full polarisation detection \citep[following the definitions
described by][]{2010PASA...27..104V}, the data were folded at the topocentric
period of the pulsar into 1024 phase bins.  The mean polarisation profile was
output every minute and a total of 25 hours of data were recorded. The
multibeam receiver is equipped with a noise diode that is coupled to the
receptors and driven with a square wave to inject a pulsed polarised reference
signal into the feed horn.  This signal was recorded for three minutes before
and after every 64 minute observation of the pulsar.  These data are archived
in, and can be obtained through, the Australia Telescope Online Archive and
CSIRO Data Access Portal\footnote{http://datanet.csiro.au/dap/}
\citep{2011arXiv1105.5746H}.

The data are stored in the {\sc PSRFITS} format and all processing was done with the
{\sc PSRCHIVE} data processing software suite \citep{2004PASA...21..302H}. First we
ensured that a recent model for the pulsar spin, astrometric and orbital
parameters \citep{2008ApJ...679..675V} was used throughout for our data
processing. To remove narrow band radio frequency interference (RFI), median
filtering was applied by comparing the total flux in each frequency channel
with that of its 49 neighbouring channels.  To avoid distortions at the edge of
the observing band, we rejected five per cent of the frequency channels on each
side of the band. A search for impulsive RFI was performed with the ``lawn
mower'' method\footnote{http://psrchive.sourceforge.net/manuals/paz }.  Pulsar
observations were calibrated for polarisation as in
\citet{2004ApJS..152..129V}. The flux density was calibrated by observing the
Hydra A radio galaxy which is assumed to have a flux of $43.1$\,Jy at
$1400$\,MHz and a spectral index of $-0.91$.

\begin{figure}
\includegraphics[angle=-90,width=\columnwidth]{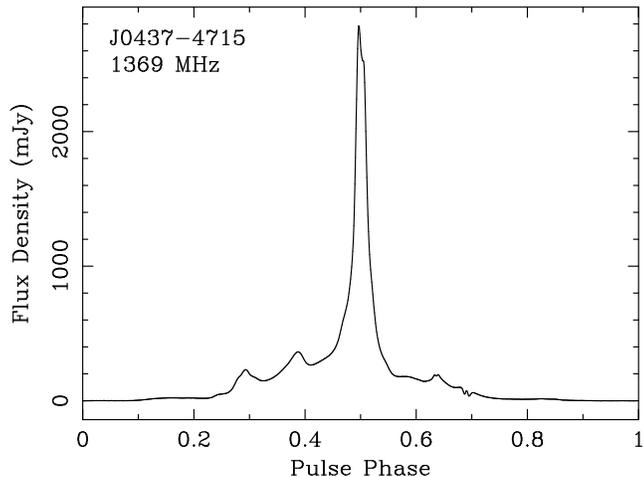}
\caption{The high $S/N$ ($15,000$) template for \psr, created from $5.5$
hours of observations. The solid black line represents the total intensity.
}
\label{Fig::std}
\end{figure}

A high $S/N$ template for each frequency channel was created by integrating the
observations obtained during the first day of data and then used to identify
and remove data affected by broadband impulsive RFI as follows.  First, for
each frequency channel, the best-fit scale, baseline offset and phase shift
\citep{1992RSPTA.341..117T} between the profile and template were applied to
compute the difference between the template and the data. Second, the rms flux
of this difference was computed after integrating over frequency at the
dispersion measure (DM) value of the pulsar and at zero DM.  RFI  will induce a
high rms flux at zero DM while, at the pulsar's DM, impulsive interference will
be smeared across multiple phase bins.  Hence, if the difference has an rms
flux value at zero DM that is higher than at the pulsar's DM the profile is
potentially polluted by RFI.  A few hundred profiles were examined by eye. The
rms ratio at both DMs for the RFI polluted difference profiles allowed the
determination of a threshold ratio above which the remaining profiles were
automatically tagged as being affected by RFI and rejected from further
analysis. After all the RFI removal stages we were left with $1145$ one minute
integrations that are considered RFI free. From the $5.5$ hours of observations
taken during the first day of observing, we created a final, frequency
integrated total intensity template shown in Figure~\ref{Fig::std} with a $S/N$
of 15,000.

\begin{figure*} 
\includegraphics[width=\textwidth]{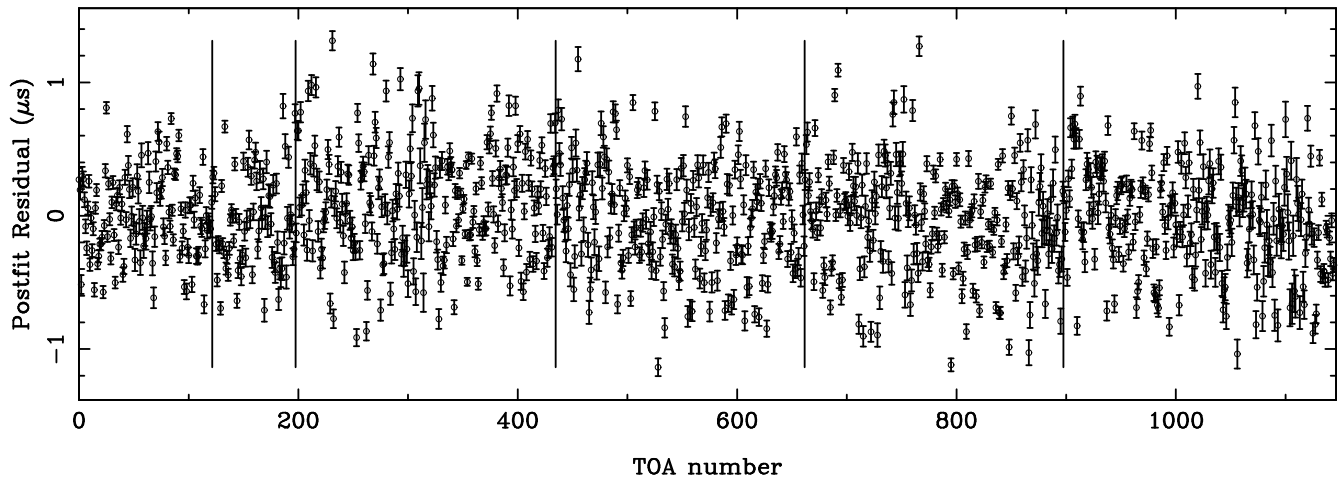}
\caption{Timing residuals for 6 days of data timed against the standard 
from Figure \ref{Fig::std}. The mean ToA estimation error is of the order of $72$~ns,
whereas the weighted rms of the residuals $\sigma_{ToA}$ is 372~ns.
The fit has $\chi/dof^2$ of 33.8. For clarity we have plotted the residuals
as a function of ToA number. The vertical lines are plotted between
observations taken on different days. \label{Fig::residuals_all_serial}}
\end{figure*}

The ToA of each observation not affected by RFI in the remaining six days was
determined by cross-correlation with the template \citep{1992RSPTA.341..117T}.
Timing residuals were formed from these ToAs and the pulsar model using the
\textsc{tempo2} software package \citep{2006MNRAS.369..655H}.  The ToA
residuals for these data are shown in Figure \ref{Fig::residuals_all_serial} as
a function of ToA number.  Note that they are not evenly spaced throughout each
day.  The mean ToA estimation error is only 72~ns and the mean $S/N$ is~$770$.
The weighted rms of the timing residuals however is $\sigma = 372\;{\rm ns}$
and the reduced chi squared of the fit, $\chi^2/dof$, where $dof$ denotes the
number of degrees of freedom, is $33.8$. The unweighted rms timing residual is
similar: $389$~ns.

The high $\chi^2/dof$ value could be caused by underestimation of the ToA
uncertainty or because the pulsar model does not accurately predict the
observed ToAs.  To verify the estimated arrival time uncertainties, we carried
out a Monte-Carlo simulation in which each observed profile is replaced by an
exact copy of the template with a suitable amount of white noise added to yield
a $S/N$ equal to that of each observation when averaged over many realisations.
The $\chi^2/dof$ of the timing residuals is always very close to unity in these
simulations, implying that the ToA uncertainties are calculated correctly under
the assumption of equation~\ref{Equation::timing}.

\begin{figure}
\includegraphics[width=\columnwidth]{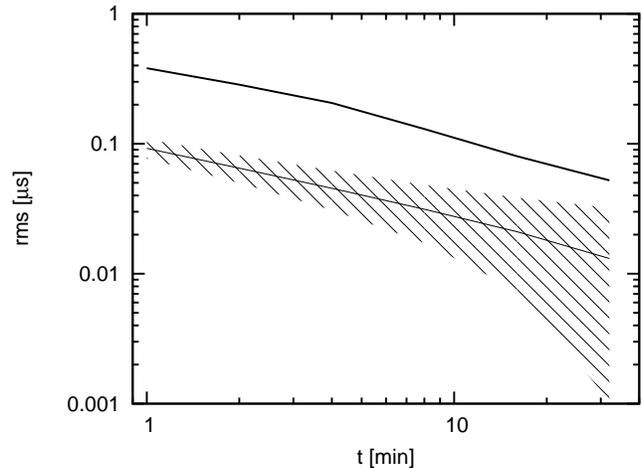}
\caption{Comparison of achieved timing precision with the theoretically
attainable precision as a function of integration time. The line in the dashed
region plots the theoretical rms, equal to the mean rms obtained from $10^5$
different realisations of white, homoscedastic noise. This values agree with
expectations based on the width of the mean pulse profile and the $S/N$. In 95\%
of the simulations, the rms falls within the dashed region. The confidence
interval is much broader at long integration lengths because a fixed number of
initial pulse profiles is used; hence, at large t, fewer independent instances
remain, thus biasing the estimate.
\label{Figure::compare}}
\end{figure}

In the above simulation, the white noise added to each simulated pulse profile
is statistically independent of the noise in every other profile; in this case,
as profiles of roughly equal $S/N$ are integrated together, the rms of arrival
times derived from the integrated totals will be roughly proportial to
$T^{-1/2}$, where T is the integration length.  Consequently, the statistical
independence of errors in arrival time estimates is commonly verified using a
plot of residual rms as a function of integration length, as shown in Figure
\ref{Figure::compare}.  Here, the thick line indicates the mean theoretical
expectation based on $10^5$ simulations of $64$ ToAs derived from template with
white noise added.  The shaded region shows the $95\%$ confidence levels
derived from the same simulations. The deviation of solid lines from
$1/\sqrt{t}$ behaviour is solely due to the small number of points at longer
integrations, e.g. only 2 points are available at 32 minute integration length.
The fact that the observed rms follows the expected behaviour suggests that no
pulse-to-pulse correlations or anti-correlations are important in our dataset.
We note that the precision measured is far worse than the theoretical
expectation. For example, with 32 minute integrations, we expect an rms timing
residual of $\sim13$~ns but we observe a typical value of $52$~ns.  This factor
of $\sim4$ worse than the theoretical prediction implies that, if this problem
was understood and fully corrected, then the same observing precision could be
achieved with integrations 16 times shorter than currently required.

\begin{figure}
\includegraphics[angle=-90,width=\columnwidth]{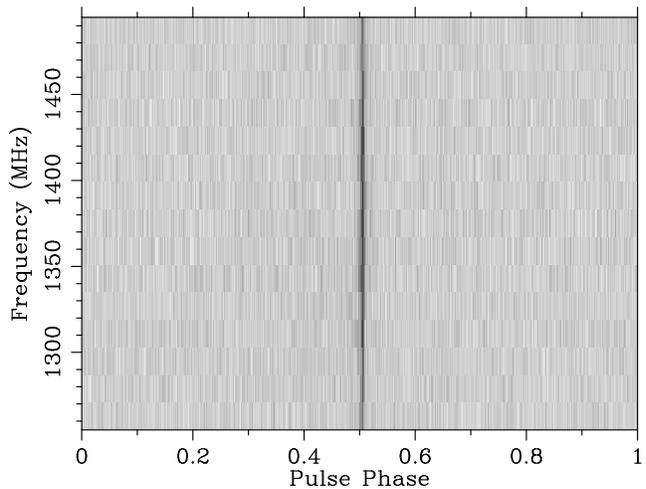}
\caption{Greyscale image of a single pulse from \psr\ as a function of pulse
phase and frequency taken with the APSR instrument. We stress that (a) the
subpulse persists across the whole available band, and (b) that each subpulse
is very different from the average profile.
\label{Figure::single_pulse}}
\end{figure}

\begin{figure}
\includegraphics[width=\columnwidth]{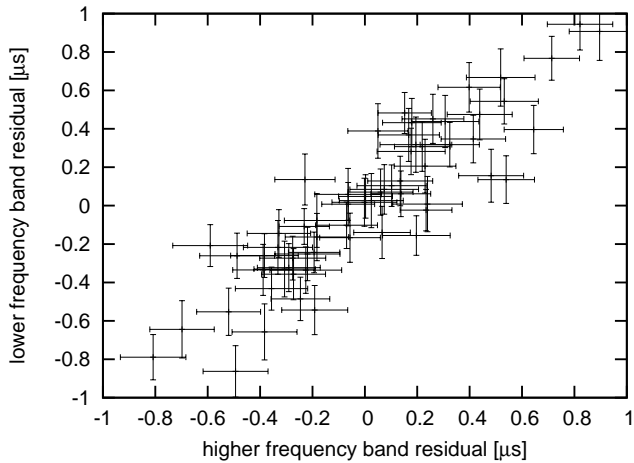}
\caption{Timing residuals derived from two independent sub-bands every 60
seconds plotted against each other to demonstrate their remarkable degree of
correlation. \label{Fig::freq_dep_against}}
\end{figure}

The above simulation does not include any self-noise  or subpulse modulation;
therefore, the predictions in Figure \ref{Figure::compare} are those expected
from the radiometer equation \citep[e.g.][]{1985ApJ...294L..25D}. However, in
the real data the variance of the noise in the off-pulse region understimates
the variance and completely neglects the temporal correlation of the noise in
the on-pulse region; that is, the actual noise is heteroscedastic and
correlated.  Increasing the gain of the antenna will amplify SWIMS and while
the $S/N$ calculated using only the off-pulse noise will increase the rms
timing residual will not decrease. Therefore, longer integrations will be
necessary to achieve a lower rms timing residual. More importantly, the SWIMS
in the total intensity from subpulse modulation is spectrally correlated.
Figure \ref{Figure::single_pulse} shows a greyscale image of a single pulse
from \psr\ as function of pulse phase and radio frequency taken with the ATNF
Parkes Swinburne Recorder \citep[APSR;][]{2011PASA...28....1V}. The emission
clearly extends across the entire observed band, producing a high degree of
spectral correlation of subpulse intensity fluctuations.  To see if ToAs from
independent bands are correlated we divided our template and each one minute
observation into two independent frequency channels and determined the timing
residuals for each channel separately.  Figure \ref{Fig::freq_dep_against}
shows the timing residuals plotted against each other for one hour of data
processed in this way and shows that the ToAs are highly correlated between the
two channels (the average Pearson product moment correlation coefficient
between the two sub-bands is $0.91$). If the subpulse modulation contribution
to SWIMS had no impact on the timing residual, no such correlation would be
present. In addition, the rms timing residual of each of the two sub-bands, as
well as the combined rms timing residual is similar to the value obtained for
the combined data.  Under the assumption that the broadband subpulse
properties of \psr\ are responsible for the scatter in the timing residuals,
increasing the observing bandwidth will not reduce the SWIMS component due to
subpulse modulation.

The heteroscedastic and both temporally and spectrally correlated properties of
SWIMS can lead to significant statistical bias in arrival time estimates
derived from sources with amplified flux densities comparable to the system
equivalent flux density.  The following sections report on an investigation of
one possible method of correcting these biases.

\section{Method}
\label{Section::method}

As explained in the previous section, failure to account for the statistical
characteristics of SWIMS leads to measurment bias and underestimation of
arrival time uncertainty.  In this section, we explore the use of principal
component analysis (PCA) to correct the statistical bias through a series of
simplified simulations.  These simulations do not model the large number of
impulsive intensity fluctuations; rather, the simulations demonstrate that the
PCA model corrects only the arrival time bias due to profile shape variations
and that no other sources of phase noise are incorrectly mitigated. An overview
of PCA is given in \citet{PCA}. We extend the analysis introduced by
\cite{2007PhDT........14D} and then present a number of illustrative
simulations that demonstrate the validity of our method and its implementation.
A very similar methodology has been under development by Cordes and his
collaborators since the 1990s \citep[private
communication,][]{1993ASPC...36...43C}.

The PCA method provides a rigourous and unbiased statistical method for
analysing temporally correlated variations in total intensity.  For each
one-minute observation of \psr\, $\sim 15\times10^6$ samples are integrated in
each of the $N_{\rm bin}$ pulse phase bins; therefore, by the central limit
theorem, the fluctuations in total intensity are well described by a
multivariate normal distribution. If the distribution of these fluctuations was
strongly non-normal, better performance might be achieved by a similar method
using independent component analysis \citep{PCA}. We note that the number of
pulses integrated in each minute is approximately an order of magnitude larger
than the number of pulses considered in previous studies of profile stability
\citep{1975ApJ...198..661H,1995ApJ...452..814R}.

Assume that $N$ such observations have been made of a given pulsar.  We
describe the profile for the $i$'th observation as a column vector\footnote{Our
notation is defined as follows: All matrices are denoted by bold sans serif
font (e.g. $\mathbfss{M}$).  All vectors are denoted by bold italic font (e.g.
\textbfit{v}).  An element of a matrix is denoted by $M_{ij}$. An $i$'th column
of a matrix is denoted by $\mathbfss{M}_i$.  An $i$'th element of a vector is
denoted by $v_i$. If there are multiple vectors of a given type, we denote the
$i$'th vector by superscripting, e.g. $\mathbfit{v}^i$. The indices $i$ and $j$
always are in range $[1,N]$ and $[1,N_{\rm bin}]$, respectively.}
$\textbfit{p}^i$. The $j$'th element, $p_j^i$, is the amplitude of the $j$'th
bin in the $i$'th profile.  The covariance matrix is typically computed after
subtracting the mean of all observations from each observation.  Here, we
assume that the template, \textbfit{s} , is a good estimate of the mean profile
and, before subtraction, each observation is first adjusted to match the
template using the best-fit phase shift, scale and offset as derived from the
template-matching procedure used for pulsar timing (see
\S\ref{Section::observations}).  We then form the covariance matrix of the
dataset by computing the outer product of template matched profiles:
\begin{equation}\label{Equation::covariance}
\mathbfss{C} = \frac{\displaystyle\sum_{i=1}^Nw_i
	\left(\mathbfit{p}^{\prime i} - \mathbfit{s}\right)\left(\mathbfit{p}^{\prime i} - \mathbfit{s}\right)^T}
	{\displaystyle\sum_{i=1}^Nw_i}\;{\rm,}
\end{equation}
where $w_i$ is the $S/N$ of the $i$'th profile, the T superscript denotes
transposition and the prime superscript signifies that the profiles have been
matched to the template. The resulting covariance matrix, $\mathbfss{C}$, is a
symmetric matrix with the number of rows and columns equal to the number of
bins, $N_{\rm bin}$, in each profile. We note that at least $N_{\rm bin}$
observations are necessary for $\mathbfss{C}$ to have full rank. Furthermore, the
data set should be large enough so that all potential modes of profile
variation are represented.

Template matching before subtracting the standard profile removes three degrees
of freedom from the shape fluctuations that are intrinsic to the pulsar signal.
For example, to first order, the best-fit phase shift removes all variations
that correlate with the derivative of the standard profile with respect to
pulse phase.  Removing variations with a certain profile shape is equivalent to
projecting the \mbox{$N_{\rm bin}$-dimensional} vector space of the total
intensity fluctuations onto the $N_{\rm bin}-$1-dimensional subspace that is
orthogonal to the axis defined by that profile shape.  A significant amount of
the fluctuation information may be lost by this projection.  However, if the
best-fit phase shift were not first removed, any actual phase shifts would be
misinterpreted as shape variations; therefore, this dimension must be excluded
from the analysis.  Similarly, the best-fit scale and offset remove variations
in pulsar flux and system temperature, respectively; these fluctuations are not
the focus of this work. In practice, all the data are fit for phase shift, flux
scale and baseline offset and these three dimensions are always projected out
of the available vector space in which the pulse profiles are described. The
eigenvectors corresponding to these three dimensions all have the same
eigenvalue (zero) and hence together form an eigenspace; we refer to the
template matching eigenspace as the fit-space throughout the remainder of the
paper.

To characterise the remaining fluctuations of the total intensity, we solve the
eigenproblem of the covariance matrix. The eigenvectors $\mathbfit{e}^j$ define the
principal axes in the \mbox{$N_{\rm bin}$-dimensional} vector space of profile
shape variations along which the intensity fluctuations are correlated as a
function of pulse phase.  Sorting the eigenvectors in order of decreasing
eigenvalue, $\lambda$, allows us to determine the most significant variations.
The variance corresponding to each eigenvector is equal to the corresponding
eigenvalue.

The eigenvectors form an orthonormal basis onto which each residual difference
profile can be projected.  The projection coefficient, $\alpha_{ij}$, of the
$i$'th difference profile onto $j$'th eigenvector is
\begin{equation}\label{Equation::projection}
\alpha_{ij} = \left(\mathbfit{p}^i - \mathbfit{s}\right)^T\mathbfit{e}^j\;{\rm .}
\end{equation}
These coefficients can be thought of as the residual of the $i$'th pulse
profile in the basis spanned by the eigenvectors $\mathbfit{e}$ and are often
referred to as the principal components.

After the subspace projection that removes the phase shift, scale, and offset
dimensions, the remaining projection coefficients for each residual profile are
uncorrelated; i.e.,
\begin{equation}
\sum_{i=1}^N \alpha_{ij}\alpha_{ik} \propto \delta_{jk},
\end{equation}
where $\delta_{jk}$ is the Kronecker delta.  However, these coefficients may
possibly be correlated with unobservable variations in the three dof that have
been removed.  These correlations are exploited by the technique developed in
\S\ref{Section::multiple}, where we introduce a new method for using these
projection coefficients to correct the timing residuals for the statistical
bias introduced by pulse shape variability. Simulations to confirm our
algorithm are presented in \S\ref{Section::simulation}.

\subsection{Correcting the timing residuals: multiple regression}
\label{Section::multiple}

\citet{2007PhDT........14D} measured the correlation between the first
projection coefficient (corresponding to the largest variance in the data) and
the arrival time residuals.  This was subsequently used to detect corrupted
data and he has shown that it could be used to remove their deleterious effect
on the timing residuals.  However, his method only used the information stored
in the first projection coefficient.  Here we apply a multi-variate statistics
method of multiple regression to simultaneously remove the effects of multiple
varying components. 

To predict the statistical bias in ToA estimate, $I_i$, caused by SWIMS we
assume that there is a linear function relating this bias to the projection
coefficients. We use the observed timing residuals, $R_i$, to determine the
best-fit parameters for this function.

We wish to predict $I_i$ using the linear predictor
\begin{equation}
\label{Equation::predictor}
I_i = a +\mathbfit{A}^{\balpha}_i
\end{equation} 
where $a$ and $\mathbfit{A}$ are the regression coefficients of the  linear
predictor.  These are determined from the observed residuals by minimising the
mean squared error between the predicted and observed residuals: $\sum_i
\left(R_i - I_i\right)^2$.  The analytic solution is given by
\citep{johnson02}:
\begin{equation}
\mathbfit{A} = \mathbfss{D}^{-1}{\bm{\gamma}}\;{\rm ,}
\end{equation}
and
\begin{equation}
a = \nu + \mathbfit{A}^T{\bmu}\;{\rm ,}
\end{equation}
where
\begin{equation}
\mathbfss{D} =  \frac{\displaystyle\sum_{i=1}^{N}
	\left(\balpha_i - <\balpha>_i\right)
	\left(\balpha_i - <\balpha>_i\right)^T}
	{N}\;{\rm ,}
\end{equation}
is the covariance matrix of the projection coefficients, $\bgamma$ is the
vector of covariances between the residuals and the projection coefficients
\begin{equation}
\bgamma_j=\left(\left(\balpha^T\right)_j - 
<\left(\balpha^T\right)_j>\right)\left(\mathbfit{R}-\bnu\right)\;{\rm ,}
\end{equation}
and $\bnu$ is a vector with each element equal to the mean of the observed
residuals. The elements $\mu_i$ are the mean values of $\balpha_i$ and
$<.>$ denotes average of vectors.  The values of $\mathbfit{A}$ and $a$ allow us to
predict the bias in ToA estimation induced by pulse shape variations.  This
predictor has minimum mean squared error and maximum correlation with the $R_i$.
Subtracting the predicted bias from the estimated arrival times has the
potential to reduce the post-fit arrival time residual rms.

The expected improvement in timing residuals can be calculated from the
projection coefficients and the observed residuals as  
\begin{equation}
\label{Eq::predicted_improvement}
\frac{\sigma^\prime}{\sigma} = \sqrt{1-\rho^2}\;
{\rm ,}
\end{equation}
where 
\begin{equation}
\rho = \sqrt{\frac{{\bgamma}^T\mathbfss{D}^{-1}{\bgamma}}{\sigma^2}}.
\end{equation}
Here $\sigma^\prime$ is the rms timing residual for ToAs with the bias removed
and $\rho$ is called the population multiple correlation coefficient. 

It is necessary to restrict the number of eigenvectors to model only pulse
variability.  Many approaches have been put forward in the literature for
determining the number of significant principal axes \citep{johnson02}.  We
introduce a new parameter, $\xi_j$, which is the Pearson's product moment
correlation coefficient between the the timing residuals, $\mathbfit{R}$, and the
projection coefficients onto the $j$'th eigenvector, that is the $j$'th row of
$\balpha$.  The standard deviation of non-significant $\xi$ values is
determined.  Significant values are identified using a Tukey's bi-weighting
scheme starting with an initial guess of the standard deviation obtained from
the median absolute deviation. The resulting standard deviation of the $\xi$
values is a robust and resistant estimator. For more details see
\citet{andrews1972} and \citet{ureda}. Starting from the last correlation
coefficient we search for three consecutive $\xi$ values that are more than
three times this measured standard deviation.  The number of the eigenvectors
used in all subsequent processing is equal to the index of last of these three
values, when counting from one. For cases in which fewer than three $\xi$
values are significant, we compare the results of using only one eigenvector
with those of using the first five.

An implementation of this method is publicly available as a part of the
{\sc PSRCHIVE} suite. The relevant application is called ``psrpca'' and it requires
the GNU Scientific Library\footnote{http://www.gnu.org/software/gsl/} to work.

\subsection{Simulations}
\label{Section::simulation}

To confirm that our method correctly detects pulse shape variations that can be
used to correct statistical bias in ToA estimation, we carry out three
simulations of data with noise in the third regime (i.e., SWIMS).  In every
simulation, each observed profile is replaced by a copy of the template profile
plus white noise and additional varying components.  The amount of white noise
is set such that in many random realisations of the simulated observation the
mean $S/N$ of the simulated profile would match that of the given observation.
Although there can be several subpulses per pulse period, we illustrate the
technique using a simple model in which only a single subpulse is added per
minute of observation. Note that this procedure does not affect the observing
parameters (such as frequency, observation time etc.) which are held fixed at
the values in the actual observations.  The resulting simulated observations
are cross-correlated with the template to form ToAs (and hence residuals) in
exactly the same manner as the actual observations.

We initially tested the trivial cases of simulated data with white radiometer
noise only with or without arbitrary phase shifts applied to the data. As
expected, no significant eigenvectors were detected in either case.  Attempting
to correct the residuals regardless of that yields no significant improvement
in the rms timing residual. These two cases demonstrate that our method will
not artificially decrease the arrival time residual arising from white noise or
arbitrary phase shifts. The latter could arise from any phase shift such as
that due to a gravitational wave and it is important that such signal is not
removed by the PCA.

\subsubsection{Simulation 1: Single, fixed component}

\begin{figure}
\includegraphics[width=\columnwidth]{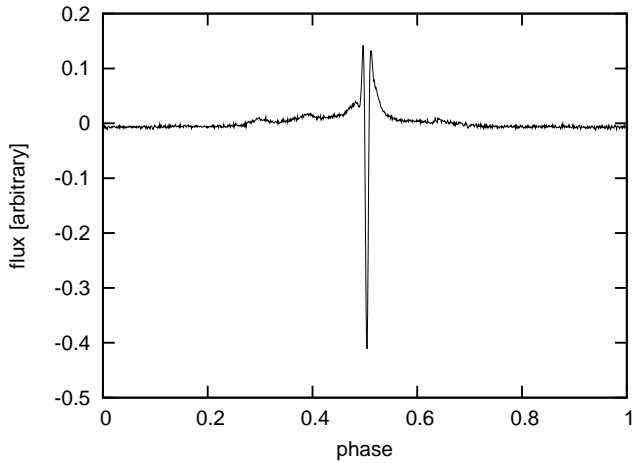}
\caption{First eigenvector for simulation with white noise, arbitrary shifts
and a von Misses component on top of the template profile. Notice that this
eigenvector does not look purely like von Mises distribution as it has to be
orthogonal to the fit-space.}
\label{Fig::pca_wn_sh_evec}
\end{figure}

Pulsar emission is often modelled as consisting of multiple Gaussian components
\citep[e.g.][]{1999ApJ...520..324K}.  Many pulsars show mode changing where one
or more components are active for only a finite amount of time
\citep{2007MNRAS.377.1383W}.  To verify that the bias introduced by a single
``mode changing'' component can be detected and corrected, we now include an
extra component in the profile that varies in amplitude.  This component is
created from a von Mises function (which is a periodic analog to a Gaussian
distribution) with a normally distributed amplitude that has a mean value of
zero and an rms of 3\% of the peak template flux.  This component is centred at
pulse phase $0.504$ and has a concentration parameter equal $0.113 {\rm
bins}^{-2}$. The resulting weighted rms timing residual was $255$\,ns and
$\chi^2/dof=14.4$, while without the additional component the same realisation
of white noise leads to an rms of $59$\,ns and a $\chi^2/dof$ of $1.0$. We
emphasise that this increased rms residual and high $\chi^2/dof$ is due solely
to the pulse shape variations and therefore should be detected and corrected
using our method.

We obtain a significant first eigenvalue.  The automatic determination of
useful eigenvectors fails, as there is only one significant eigenvector,
associated with the introduced profile variation. Figure
\ref{Fig::pca_wn_sh_evec} shows the first eigenvector, which is different from
the introduced component as it is projected onto the $N_{\rm bin}-3$
dimensional space, in which the three dimensions corresponding to the fit-space
are removed.  Although these degrees of freedom have been removed, the residual
shape variation is still highly correlated with the ToA residual, as discussed
above.  Correcting the residuals using the one significant eigenvector reduces
the rms residual to $67$\,ns and the $\chi^2/dof$ to $1.0$. We note that this
improvement agrees very well with prediction from equation
\ref{Eq::predicted_improvement}. We calculate reference residuals by simulating
and timing data without any additional components but with the same realisation
of white noise as in the simulation with the varying component. As expected the
corrected residuals are highly correlated with the reference residuals
(correlation coefficient of $0.93$). This indicates that we have recovered most
of the signal in the original residuals and have removed the effect of the
pulse variability.

We note that after removing the bias in ToA residuals the  weighted rms timing
residual is  almost the same as in the case of reference residuals; however,
the $\chi^2/dof$ can be smaller than that of the reference residuals because
the ToA uncertainties have increased. The total intensity fluctuations decrease
the cross-correlation between the observation and the template, thereby
increasing the estimated uncertainty \citep[see equation A10
from][]{1992RSPTA.341..117T}.

In order to check the effect of using a different number of eigenvectors to
correct the timing residuals, we re-analysed the data using five eigenvectors.
In this case the rms residual and $\chi^2/dof$ were $64$\,ns and $0.92$
respectively. The corrected residuals are still pleasingly highly correlated
with the reference residuals. We therefore conclude that our result is not
highly sensitive to the number of eigenvectors used, but care needs to be taken
when choosing that number.

In addition, we also tested an extended version of this simulation, where
arbitrary phase shifts were included to simulate, e.g., the effect of
gravitational waves. As expected, only the bias in the timing residuals due to
the introduced profile variation is removed and the arbitrary phase shifts
remain unaffected. Again, the residuals are highly correlated with reference
residuals where the same realisations of white noise and arbitrary shifts were
introduced.

\subsubsection{Simulation 2: Multiple, fixed components}

\begin{table}
\caption{Parameters of the multi-component simulation.}
\begin{tabular}{ c | c  c  c  c  c  c }
		&	1	&	2	&	3	&	4	&	5	&	6	\\
		\hline
centre	& $0.504$	& $0.496$	& $0.512$	& $0.524$	& $0.509$	& $0.520$	\\	 
	 
amplitude	& $0.74\%$	& $0.36\%	$ & $0.86\%$	& $1.4\%$	& $0.7\%$ & 
$0.4\%$\\
$\kappa$ & $0.452$	& $0.164$ & $0.098$ & $0.050$ & $0.577$ & $0.104$ \\
\end{tabular}
\label{Tab::gauss_multi}
\end{table}

The previous simulation dealt with only a single varying component while many
pulsars can emit radiation simultaneously from several components.  Even if
only one of the multiple components was present in each rotation of the pulsar,
several components would be present after integration over multiple pulse
periods.  To demonstrate again that PCA and multiple regression do not remove
phase shifts that are not caused by pulse shape variations, we now introduce
six von Mises functions whose amplitudes are allowed to vary independently. The
parameters of these components are presented in Table \ref{Tab::gauss_multi},
where centre is the central phase of a von Mises component in pulsar turns;
amplitude is the rms of the amplitude distribution of given component, in units
of the template's peak flux; and $\kappa$ is the concentration parameter of the
von Mises distribution in units of ${\rm bin}^{-2}$. We also apply arbitrary phase
shifts after adding the varying components. This leads to a weighted rms
residual of $372$~ns and $\chi^2/dof$ of $32.1$. Note that the reference
residuals obtained from a simulation with the same realisation of white noise
and arbitrary phase shifts, but no additional components, have an rms of $206$
ns  and $\chi^2/dof$ of $12.2$ due to the arbitrary  phase shifts; i.e., we do
not expect the rms to be of the same order as the ToA measurement error after
bias removal.

Our method gives corrected residuals with weighted rms residual of $216$~ns
($\chi^2/dof=10.8$) and they are highly correlated with the reference
residuals. With multiple components the bias is not completely removed because
more than one projection coefficient correlates with the arrival time residual
and these projection coefficients may not be statistically independent of each
other.  Nevertheless, significant improvement in the rms residual and in the
$\chi^2/dof$ is apparent and much of bias is removed. The additional shifts,
corresponding to unmodelled timing noise processes are still unaffected; i.e.,
the post-correction residuals are highly correlated (correlation coefficient of
$0.87$) with the reference residuals.

\subsubsection{Simulation 3: Single, random component}

We now consider a possibly more realistic case of the pulsar emission being
erratic and distributed in phase over the whole region in which the average
profile is visible; this case corresponds to the stochastic nature
\citep{1985ApJS...59..343C,1993ASPC...36...43C} of modulated pulses. In this
simulation we allow a single von Mises component per simulated profile to be
centred anywhere in within the central peak (central phase uniformly
distributed between $0.479$ and $0.518$). The concentration parameter of the
component is uniformly distributed between $0.055$ and $1.386$ ${\rm bin}^{-2}$. The
amplitude of this component is normally distributed with an rms equal to
$2.3\%$ of the intensity of the template at the centre phase of the component.
This value is chosen in order that the resulting rms timing residual is
$380$~ns and $\chi^2/dof$ value of $35 .1$, both very close to the observed
values for \psr.

Multiple significant eigenvectors are detected, with the number varying between
different realisations from 10 to 50. Correcting the residuals reduces the rms
to $266$~ns and $\chi^2/dof$ to $17.2$. We note that in this simulation no
arbitrary shifts were included and therefore we conclude that the PCA method
has failed to completely remove the bias in ToAs (and hence in residuals)
induced by this kind of erratic shape variation.  As explained in
\S\ref{Section::method} some fraction of the fluctuation power is lost during
the vector subspace projection effected by removing the best-fit shift, scale,
and offset. As before, for data integrated over multiple pulse periods, more
than one varying component per pulse profile would be present and this could
make it even more difficult to remove the bias in ToAs.

\section{Results}
\label{Section::results}

\begin{figure}
\includegraphics[width=\columnwidth]{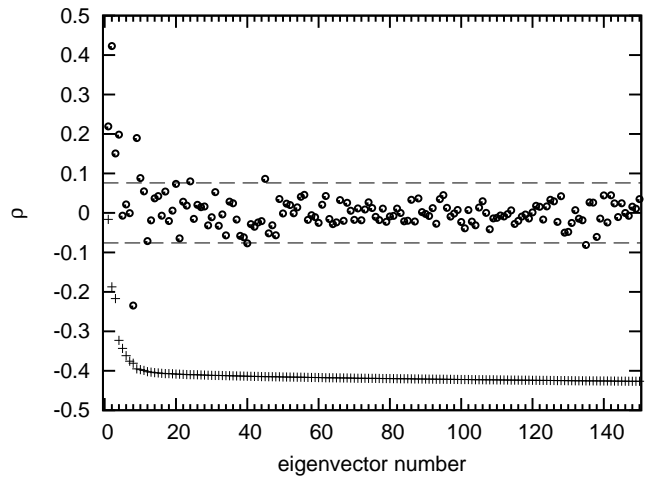}
\caption{Distribution of correlation coefficients  between the residuals and
projection onto eigenvectors for the actual observations, shown by open
circles.  Several correlation  coefficients are more than 3 standard deviations
(denoted by the dashed lines) above the background noise level. Only the first
150 correlation coefficients are shown, the rest of the correlation
coefficients looking very similar as the last shown. The crosses are showing
the corresponding eingenvalues normalised by their sum, multiplied by a factor
of 10 and offset by $-0.45$ for clarity of the figure.
\label{Fig::pca_real_corel}}
\end{figure}

Applying our method to the observed dataset of 1145 profiles leads to:
\begin{itemize}
\item the detection of significant pulse shape variations with at least ten
significant eigenvectors,
\item a reduction in rms timing residual from $372$~ns to $294$~ns and a
reduction in $\chi^2/dof$ from 33.8 to 21.1. 
\end{itemize}

\begin{figure*}
\includegraphics[width=\textwidth]{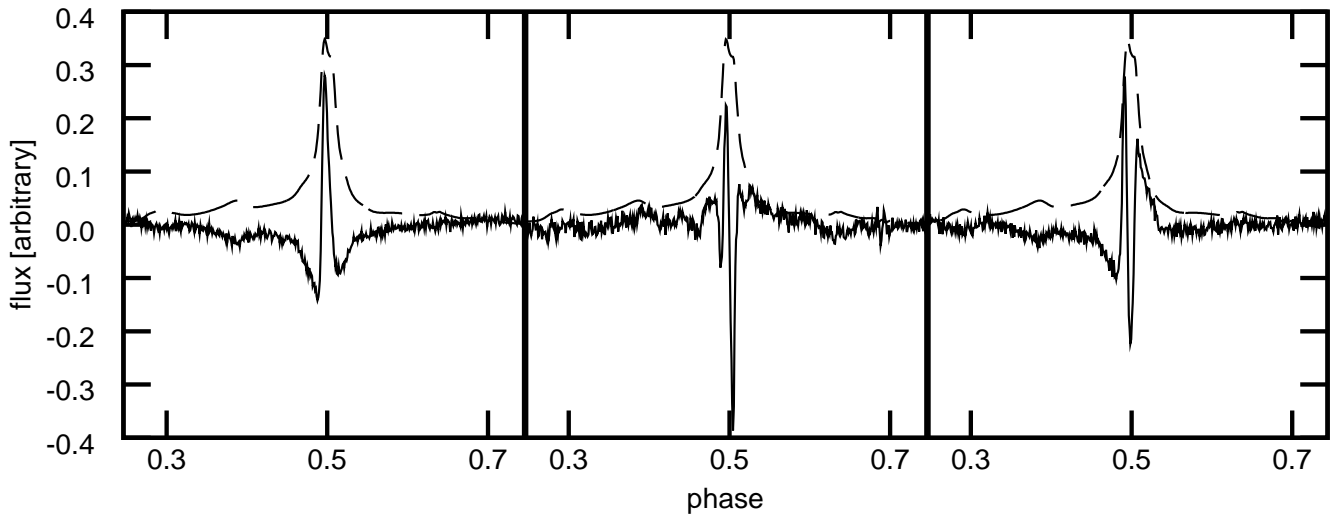}
\caption{First three eigenvectors for the observed data. We show only the
central part of the profile for clarity. The vertical thick lines separate the
eigenvectors. The dashed line represents the scaled template added for
reference.
\label{Fig::pca_real_evec}}
\end{figure*}

The first 150 values of $\xi$ are plotted in Figure~\ref{Fig::pca_real_corel}
which indicates that around 10 eigenvectors significantly correlate with the
observed residuals while the rest are consistent with being white noise. The
choice of exact number of eigenvectors would be difficult to make without the
rigourous criteria described before. In Figure \ref{Fig::pca_real_evec}, we
present the three most significant eigenvectors overlaid on the pulsar template
profile.  The detection of significant eigenvalue-eigenvector pairs is a direct
consequence of temporally correlated fluctuations in total intensity; i.e.,
significant shape variations are detected.  As discussed in \S
\ref{Section::intro}, \S\ref{Section::selfnoise}, and \S
\ref{Section::observations}, it is likely that such variations originate from
the pulse-to-pulse variations of the pulsar emission. The most significant
variation occurs at phase $0.497$, in the main peak of the pulse profile. The
majority of other significant eigenvectors peak around the main and second
highest peak in the mean profile. One exception is the 10th eigenvector that
peaks at phase $0.385$, that is in the local peak on the left hand side of the
main peak.

Using the 10 most significant eigenvectors to correct the bias in arrival time
estimates, the rms timing residual was reduced by $20\%$ and $\chi^2/dof$ was
reduced by $36\%$. Only $3\%$ of the variance in the timing residuals can be
attributed to the most significant eigenvector; therefore, multiple regression
is required to provide the best estimate of statistical bias. The third
simulation has demonstrated that pulse profile variability can introduce bias
in ToAs that cannot be removed completely using our method and the timing
residuals after bias correction are still biased. Therefore, even though only
$20\%$ of the rms timing residual is corrected, there need not be another
explanation for the remaining timing noise. For completeness, other effects
that might contribute to the rms of timing residuals are discussed in the next
section.

To investigate if the intensity fluctuations arise from a stationary stochastic
process, at least in the wide sense, we used five hours of observations of
\psr\ during July 2009 obtained using the same observing system as described in
\S\ref{Section::observations}. These additional data were processed to form
pulse profiles, arrival times and timing residuals in exactly the same way as
our main data set.  We used the eigenvectors and regression coefficients
obtained above to correct these July 2009 timing residuals.  We again achieved
a 22\% decrease in the rms timing residual from $380$\,ns to $296$\,ns and the
$\chi^2/dof$ was reduced by $36$\% from $39.1$ to $24.9$.  The fractional
improvement is similar to before, implying that the covariance matrix and hence
profile variability is stationary in time and that after the regression
coefficients and eigenvectors have been determined for one dataset they can
subsequently be applied to other observations of the same pulsar obtained with
the same instrumentation and observing parameters.

\section{Discussion}
\label{Section::discussion}

We first discuss the method that we have introduced to search for and correct
pulse shape variability.  Second, we discuss the astrophysical implications of
broad-band profile shape variations intrinsic to the pulsar.

\subsection{Discussion of the method}

It is clear from the simulations that our method successfully detects
significant pulse shape variations and partially corrects the bias induced in
timing residuals due to such variations in many cases.  The method has certain
limitations. First, in order for the covariance matrix, $\mathbfss{C}$, to have full
rank, the method requires at least $N_{\rm bin}$ observations. With fewer
observations than phase bins, it may be necessary to average adjacent phase
bins, or, if prior knowledge suggests that the pulse variability occurs in only
a restricted region then only these bins could be included in the analysis. If
full phase resolution is required then the covariance matrix can be determined,
but not all of the eigenvectors can be calculated. Alternatively, following
\citet{2007PhDT........14D}, the PCA method can be developed in the frequency
domain using only the significant harmonics thus reducing the number of
required observations.

A large number of observations is desirable for two reasons. Firstly, dividing
a fixed observing time into smaller intervals provides a greater number of
estimates of the pulse profile, thereby increasing the $S/N$ of the covariance
matrix estimate. Secondly, even for $N>N_{\rm bin}$ our method is limited by
the SEFD present in all observations.  Such noise will reduce the precision
with which the eigenvectors may be determined, reducing their ability to fully
describe the shape variations. For instance, in the timing residuals of
simulations with only white radiometer noise added, even though no significant
eigenvalues were measured, the rms timing residual is reduced by a negligible
amount when using just a few eigenvectors.  When all of the eigenvectors are
included in the bias removal, the effects of white noise are artificially
reduced. This occurs because the white radiometer noise affects the regression
coefficients $a$ and ${\mathbfit{A}}$ in equation \ref{Equation::predictor}. Since
the eigenvectors are measured using the data that are being ``corrected', the
noise in the eigenvectors correlates with the noise in the data. This is
similar to the ``self-standarding'' effect that arises when the mean of a set
of observed profiles is used as the template to derive arrival times from the
same data, as described in Appendix A of \citet{2005MNRAS.362.1267H}. Applying
the eigenvectors to a completely independent data set leads to no significant
change to the timing residuals as the noise in the data no longer correlates
with noise in the eigenvectors.  The degree of correlation between individual
residual profiles and the white-noise eigenvectors may also be reduced by
increasing the number of observations from which the covariance matrix
$\mathbfss{C}$ is estimated.  If the eigenvectors are obtained from the data to be
corrected, then it is essential to apply the rigourous criteria described in \S
\ref{Section::multiple} to choose only the significant eigenvectors. 

Many pulsar observations are affected by radio frequency interference (RFI) and
the measured eigenvectors are extremely sensitive to the presence of RFI in the
data. As pointed out by \citet{2007PhDT........14D}, PCA is also a sensitive
and robust method for detecting RFI and other types of data corruption and
distortions. It is therefore essential that the data used in forming the
eigenvectors are unaffected by RFI.  In our case nearly $32\%$ of the pulse
profiles had to be rejected. RFI might also be mitigated through the use of a
robust estimator of the covariance matrix, such as the minimum covariance
determinant estimator \citep{WICS:WICS61}.

In contrast to more traditional applications of the PCA method, our method
relies on first aligning each pulse profile to the template using the best-fit
phase shift, scale, and baseline offset.  As a result of this fit, the last
three eigenvalues are several orders of magnitude smaller (i.e., close to zero)
and the corresponding eigenvectors spanning the fit-space are highly correlated
with the template profile, its phase derivative, and the baseline offset or
their linear combinations.  In other words, the profiles are originally
described in an $N_{\rm bin}$ dimensional vector space and the fit projects the
data onto $N_{\rm bin}-3$ dimensional vector space in which the fit-space has
been removed. This removes three degrees of freedom from the remaining
eigenvectors and limits the efficacy of the correction scheme presented in this
paper because the fit-space component of the intrinsic shape fluctuations has
been removed.  Consequently, the bias due to any intrinsic shape fluctuations
that correlate with these eigenvectors cannot be fully corrected. This is
especially important in the case of profile variations correlated with the
template derivative as such variations will introduce most bias in the ToAs.
Only the total intensity fluctuations that are orthogonal to the fit-space
contribute to the predictor computed in equation \ref{Equation::predictor}.
The degree to which such fluctuations are correctable depends on how strongly
they correlate with these three eigenvectors and the degree of correlation
between the remaining projection coefficients and the arrival time residual.

We note that our work has implications for any relevant template matching
algorithms. Such algorithms normally assume that the errors in the measurements
of intensity are homoscedastic and uncorrelated. Our detection of profile
variations implies that the errors are, in fact, heteroscedastic and
correlated. We note that including the covariance matrix, which carries the
information about the correlation and heteroscedasticity of the noise, into the
template matching algorithms will yield better estimates of ToA uncertainty. It
may also remove the necessity of correcting the residuals by the means
described in this paper because  any statistical biases caused by the intensity
fluctuations might be removed at the time of ToA determination. This is similar
to the Cholesky decomposition which can remove bias in estimation of various
parameters, such as parallax, when estimating from residuals with red noise
present \citep{2011MNRAS_TMP_C}. Examples of template matching algorithms are
the standard ToA derivation algorithm presented by \citet{1992RSPTA.341..117T}
and the matrix template matching algorithm that allows all Stokes parameters to
be used in the ToA estimation \citep{2006ApJ...642.1004V}. An unbiased
generalisation of template matching would be a logical extension of this work.

\subsection{Application to \psr}	

Using one-minute integrations of \psr, we have detected shape fluctuations that
we attribute to the stochastic subpulse structure of the pulsar emission.
However, for many timing applications, the exact cause of the pulse shape
variations is irrelevant.  This technique can be used to correct bias and
improve sensitivity to any phenomena that do not induce shape variations.  For
instance, in order to place a limit on the existence of a gravitational wave
background \citep[e.g.][]{2006ApJ...653.1571J} some methods compare the amount
of power in the timing residuals with the power predicted to be induced by
gravitational waves.  Such waves will not affect the pulse profile shape.
Therefore, reducing the rms timing residuals by accounting for pulse shape
variations allows an improved limit on the existence of a gravitational wave
signature to be obtained. We note that our correction method does not remove
any of the signal induced by gravitational waves or any other phase shift of
the pulsar profile, as verified by the simulations.

The long term timing of \psr\ shows significant low frequency structure present
in the residuals \citep[e.g.][]{2008ApJ...679..675V}. Such red timing noise is
a common type of non-Gaussianity (in general any asymmetric distribution will
have a similar effect) in the timing residuals. It is important to determine
the best predictor for correcting the residuals based on a short data span, as
it will be less affected by any non-Gaussian noise. In the presence of
non-Gaussian components, the best predictor will be affected as the correlation
coefficients $\xi$ between the residuals and projection coefficients will be
biased toward zero.  Presence of Gaussian noise (or any other symmetrically
distributed noise) will increase the uncertainty in the estimate of $\xi$
values but will not bias them.

The timing residuals for our Parkes observations are only partially corrected
by our new method. We have demonstrated that partial correction does not
necessarily imply the existence of other sources of scatter in ToA residuals
that do not affect the pulse shape. At the same time it is not possible to
completely exclude the existence of other effects such as hardware or software
errors that are affecting the timing residuals at a lower level.  Other
possible effects that can increase the rms timing residual are described in
detail by \citet{2010arXiv1010.3785C,2011MNRAS.tmp.1442L}.  Such issues are beyond the scope of
this paper, which concentrates solely on the pulse shape variability.  We note
that if any other non-white process affecting the residuals could be corrected,
the residuals induced by profile shape variations could be removed more
completely as the estimates of $\xi$ are affected by the presence of phase
noise.

We stress that the intrinsic shape variations lead to the heteroscedastic and
correlated component of the SWIMS\footnote{We postpone the calculation of the
expected amount of ToA scatter introduced by SWIMS (based on the measured
covariance matrix) to following publications.}.  The uncorrelated component,
originating from the standard radiometer noise in the weak source limit and the
self-noise, is described by the diagonal of the covariance matrix $\mathbfss{C}$
while the temporally correlated part is described by the off diagonal elements.
However, whatever phenomena contribute to the correlated component will
naturally also contribute to the diagonal of the covariance matrix. As the
self-noise and subpulse modulation are all measured and described
simultaneously by the covariance matrix, it is natural to consider them all as
one phenomenon, which we called SWIMS throughout this paper. Two details of
covariance matrix we would like to stress are that: a) it does not contain any
information about the spectral correlation of the noise and b) the measured
covariance matrix also contains a contribution from the SEFD which can be
subtracted if necessary. Extrinsic sources can also introduce a heteroscedastic
and correlated component of noise, such as lightning strikes, other types of
RFI or some instrumental effects, such as non-linearity of the backends. The
effects of impulsive interference should be present across the whole pulse
profile as they occur randomly in pulse phase.  After careful removal of RFI,
the measured eigenvectors are consistent with white noise outside the mean
profile (see \S \ref{Section::results}).

\begin{figure}
\includegraphics[width=\columnwidth]{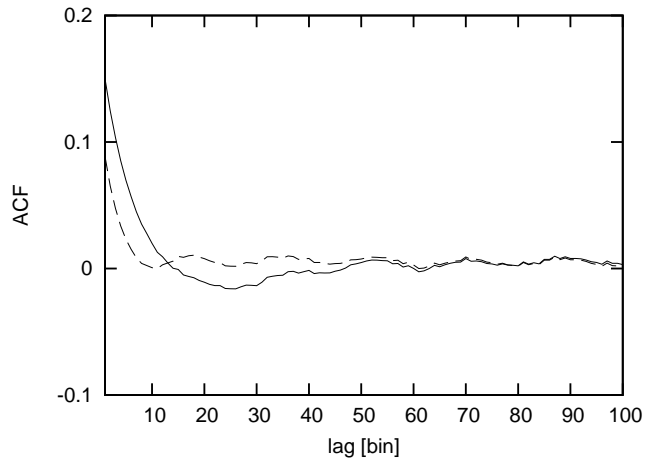}
\caption{The ACF calculated with the assumption of the timing model accurately
predicting the pulse phase over the course of observations (solid line) and
without this assumption, i.e., when performing a full template matching (dashed line).
The width of the ACF is determined by the characteristic width of single pulses
\citep{1975ApJ...197..185R}.  The SEFD contributes an unresolved spike at zero
lag, which is scaled to unit height for the solid line and the dashed line
uses the same scaling factor. }
\label{Fig::ACF}
\end{figure}

As shown in Figure \ref{Fig::pca_wn_sh_evec}, interpretation of the
eigen-vectors derived from the covariance matrix is complicated by the fact
that the fit-space has been projected out of the \mbox{$N_{\rm
bin}$-dimensional} space of the shape variations. To investigate the structure
of the shape variations prior to this projection, one can make the assumption
that the pulsar timing model accurately predicts pulse phase (at least over the
time-scale of the observations) and compute the covariance matrix of observed
profile residuals after template matching by varying only the scale and offset
(i.e. no phase shift). In this case, the covariance matrix contains the cyclic,
phase-resolved autocorrelation [autocorrelation function (ACF)] of the intensity fluctuations. The mean
ACF computed by summing elements along the diagonals of this matrix is plotted
in Figure \ref{Fig::ACF}.  The characteristic width of ACF, as determined by
fitting a Laplace function, is equal to 67 $\mu$s, which is consistent with the
average width of microstructure events reported by \citet{1998ApJ...498..365J}.
For comparison, the mean ACF formed from the covariance matrix used for bias
removal (i.e. best-fit phase shift removed) is also plotted with a dashed line.
A large fraction of the fluctuation power is removed by fitting for the phase
shift during template matching.  The phase shift fit removes variations that
correlate with the template derivative, and the autocorrelation of the
template derivative reaches its first minimum (below zero) at roughly the width
of the pulse profile (around 140 $\mu\rm{s}$).  This corresponds with the first dip seen
in the ACF formed from data in which no phase shift has been removed; no dip is
present in the ACF formed after fitting for phase shift. The measured
ACF may be affected by other non-intrinsic effects,
especially those related to propagation through interstellar medium
\citep[ISM;][]{2003A&A...405..795S}. We argue below that the ISM is not an important
factor in considerations of \psr. The ACF shows a periodic ripple (most apparent
at large phase lags), which is believed to be an instrumental artefact; it has a
period of roughly $100\; \mu\rm{s}$. Its origin is currently unknown. Preliminary analysis
of data from another instrument
\citep[Caltech Parkes Swinburne Recorder 2, CPSR2; ][]{2003ASPC..302...57B,2005PhD.....H} confirms
that this effect is intrinsic to PDFB3 as the ACF calculated from CPSR2 has
no periodic ripple present. The pulse profile variations are still detected thus confirming
their origin as intrinsic to the pulsar. 

The interstellar medium can also introduce shape variations, such as broadening
of the pulse profile, which increases  quadratically with DM and decreases
quartically with frequency. Given the very low ${\rm DM}=2.64$ cm$^{-3}$pc and
the observing wavelength $\sim 20$ cm used in this work, the expected
variations in the pulse width for \psr\ due to broadening are of the order of
$1$~ns \citep{2004ApJ...605..759B,2006A&A...453..595G}.  Interstellar
scintillation combined with frequency dependence of the pulse profile can also
lead to fluctuations of ToAs. As shown in Figure~\ref{Fig::freq_dep_against}
the residuals are highly correlated when estimating the ToAs from two separate
frequency channels.  This high correlation implies that diffractive narrow-band
interstellar scintillation is not responsible for the additional scatter.  As
we observe variations on a time-scale of minutes it is unlikely that broadband
refractive scintillation is a contributing factor to the observed pulse shape
changes as the time-scale for such variations is of the order of $1000$~s
\citep{2006A&A...453..595G}. Based on the high degree of correlation between
arrival times measured in separate bands (Figure \ref{Fig::freq_dep_against})
and the observed broadband nature of single pulses (Figure
\ref{Figure::single_pulse}), we conclude that the intensity fluctuations are
correlated over wide bandwidths.  Consequently, increasing the bandwidth does
not increase the timing precision.  Only active mitigation or longer
integrations can reduce the timing rms if the fluctuations of total intensity
are indeed the main cause of the arrival time variations that greatly exceed
the predicted uncertainty.  Simultaneous multifrequency observations of \psr\
would help to determine if the shape variations are persistent over very wide
bands. Some observations to date have shown that the giant pulses from Crab
extend over GHz bandwidths
\citep{1999ApJ...517..460S,2003Natur.422..141H,2007ApJ...670..693H}. Another
group has shown that the single pulses in \mbox{PSR B0329$+$54} persist across
$1.3$ GHz \citep{2001A&A...379..270K} but this may not be true for all
subpulse structure in the general population of pulsars.

It is also worth considering the impact of polarization variations on arrival
time estimates. Emission from \psr\ is highly polarised and a sudden change in
the position angle of the linear polarisation occurs near the peak in the total
intensity profile \citep[as noted by][]{1997ApJ...486.1019N}.  This implies
that poor polarisation calibration will lead to significant profile changes, as
quantified by \citet{2006ApJ...642.1004V}.  For a single dish that is not
equatorially mounted, these variations occur on time-scale of hours. Pulse
profile shape changes detected by \citet{1998ApJ...501..823V} were argued to be
caused by calibration errors \citep{1997ApJ...478L..95S}. Even though we detect
fluctuations of the total intensity pulse profile on much shorter time-scales,
we investigated if this was the case in our data as well. We studied the bias
in timing residuals induced by the measured pulse shape variation as a function
of parallactic angle.  We found that there was no dependence between these two
quantities. We also applied the PCA to uncalibrated data and a correlation
between the induced bias $I$ and the parallactic angle was readily apparent.

We conclude that the shape variations are more likely to originate at the
pulsar rather than in the observing hardware or from interference. The
polarisation calibration has been performed sufficiently to alleviate at least
minute-time-scale fluctuations and does not introduce detectable pulse shape
variations. The interstellar medium is also unlikely to cause such variations.
Even if the detected variability is not intrinsic, the presented methodology
remains valid. The intrinsic variation is expected from the stochastic
subpulse structure  and will be detectable if the pulsar is bright enough.

Since the detected variations are likely to be intrinsic to the pulsar, a
question arises whether the profile variations in \psr\ are related only to
SWIMS or if they are due to mode changing. We searched for clustering in the
space spanned by the projection coefficients onto the ten significant
eigenvectors by applying a friends-of-friends algorithm known from n-body
simulations to identify dark matter haloes \citep{1985ApJ...292..371D}.  We did
not find any evidence of clustering in this space and hence conclude that the
pulse profile variations are not due to mode changing.

We would like to stress the importance of our work for the next generation
telescopes, which are likely to provide more sensitivity than currently
available. With its huge collecting area of 1\,${\rm km}^2$, the Square
Kilometre Array (SKA) is expected to revolutionise pulsar astronomy. One of the
Key Science Projects of the SKA requires pulsar observations with the highest
possible timing precision \citep{2004NewAR..48..993K}. It is assumed that the
SKA will observe of the order of $100$ millisecond pulsars with an rms timing
precision better than $100$~ns.  With the SKA's phenomenal sensitivity, the
$S/N$ of a pulse profile should be $>$1000 on a time-scale of only minutes for
many pulsars (compared with many hours with the Parkes telescope). The short
observing times required to achieve such high $S/N$ ratios would allow the SKA to
observe multiple pulsars in a short time.  However, the increased sensitivity
of next-generation telescopes will also increase the relative importance of
SWIMS as the radiometer noise is decreased. If the intrinsic pulse shape
variations that we have detected for \psr\ are typical of many MSPs at the
observing frequency being used, they will induce a floor on timing precision
that can be ameliorated only with longer integration times and active
mitigation using methods such as the one presented in this work.
\citet{1985ApJS...59..343C} demonstrated that, for the majority of their sample
of 24 pulsars, timing precision is likely to be limited by phase jitter. No
millisecond pulsars were included in this sample. \citet{2010ApJ...725.1607S}
later argue that the timing noise of millisecond pulsars is similar to that of
classical pulsars, only much smaller.  Although SWIMS has not yet been detected
in most pulsars, it is likely to be revealed with better instrumentation, more
sensitive telescopes or longer data spans.  Since the detected pulse shape
variations are likely to be very broadband, increasing bandwidth will not
reduce the bias introduced by SWIMS.  This must be considered when predicting
the potential science of current and future pulsar timing array projects and
the observing time and strategy necessary to achieve the stated goals. For
example, with an SKA-like telescope, if many pulsars in a timing campaign are
limited by SWIMS, then it is better to observe multiple pulsars simultaneously
with fraction of the array for longer time rather than using full sensitivity
to observe pulsars one by one for a short time. Some proposed astrophysical
experiments demand extremely accurate ToAs over short intervals of the order of
minutes, such as when a pulsar passes behind black hole in a close binary.
SWIMS will make such experiments difficult or impossible.

The relative importance of the correlated component of SWIMS is also expected
to vary between pulsars as it depends on intrinsic subpulse emission
properties and the shape of the average pulse profile. The fractional
improvement in rms timing residual is expected to vary from case to case and it
can be hoped that for pulsars with simpler and/or narrower profiles, variability
in the subpulse structure will be less severe.  As demonstrated by our first simulation, in simple
cases our method works very well and can completely remove the statistical bias
in ToAs for some pulsars. Our method can be used to identify pulse profile
modes which can lead to improved timing.

We note that, for current telescopes, equation \ref{Equation::timing} is a good
approximation for the vast majority of MSPs, which have time averaged mean flux
densities an order of magnitude smaller than \psr.  For example, in the Parkes Pulsar Timing Array 
sample, fluxes vary between $1.3$ and $13.8$~mJy \citep[see Table 2
of][]{2011MNRAS.414.2087Y} with a median value of $2.4$~mJy, compared to the
mean value of $150$~mJy for \psr.  Consequently, the profile variability
arising from SWIMS has been neglected to date. We note that future telescopes
like the SKA are likely to perform timing array  experiments at higher
frequencies to avoid some of the problems caused by the interstellar medium.
Whether SWIMS will be a crucial limitation to precision timing for the
majority of pulsars at all observing frequencies remains to be seen.

\section{Conclusion}

\label{Section::conclusions}

We have developed an extended principal component analysis method that is
applicable to searching for pulse shape variations in pulse profiles. Applying
this method to \psr\ shows the presence of pulse profile variability that is
likely to be intrinsic to the pulsar. The statistics of this variability are
consistent over many months. The detection of significant intensity
fluctuations implies that self-noise may be a limiting factor for timing
precision of \psr\ for current generation of telescopes. Future technological
developments including construction of larger antenna and increased
instrumental bandwidth will not improve timing precision as the subpulse
structure is a source-intrinsic broadband phenomenon.  However, the effects of
SWIMS can be partially corrected by the method presented in this work and the
proposed generalised template matching.

\section*{Acknowledgements}

The Parkes Observatory is part of the Australia Telescope National Facility
which is funded by the Commonwealth of Australia for operation as a National
Facility managed by CSIRO. We thank the staff at Parkes Observatory for
technical assistance during observations. The authors are grateful for helpful
discussions with Carl Gwinn, Xavier Siemens, Ryan Shannon, Richard Manchester
and Mike Keith. We thank the anonymous referee for valuable comments that helped
improve the text. This work is supported by Australian Research Council grant \#
DP0985272. GH is supported by an Australian Research Council QEII Fellowship
(project \# DP0878388).
  
\newcommand{\apj}{ApJ}
\newcommand{\aj}{AJ}
\newcommand{\apjs}{ApJS}
\newcommand{\apjl}{ApJ Lett}
\newcommand{\nat}{Nature}
\newcommand{\aap}{A\&A}
\newcommand{\prc}{Phys.~Rev.~C}
\newcommand{\prd}{Phys.~Rev.~D}
\newcommand{\physrev}{Phys. Rev.}
\newcommand{\mnras}{MNRAS}
\newcommand{\nar}{New Astronomy Reviews}
\newcommand{\pasp}{PASP}
\newcommand{\pasj}{PASJ}
\newcommand{\apss}{ApSS}
\newcommand{\aapr}{AAPR}
\newcommand{\aaps}{A\&AS}
\newcommand{\physrep}{Phys. Rep.}
\newcommand{\sovast}{Soviet Astron.}
\newcommand{\pasa}{Publ. Astron. Soc. Aust.}

\bibliographystyle{mn2e}
\bibliography{stability}

\label{lastpage}

\end{document}